\documentstyle[aps,prb,epsf,floats,eqsecnum]{revtex}
\begin{document}
\def\SNG{{\em Physical Review Style and Notation Guide}}
\def\LUG {{\em \LaTeX{} User's Guide \& Reference Manual}}
\def\btt#1{{\tt$\backslash$\string#1}}%
\def\REVTeX{REV\TeX}
\def\AmS{{\protect\the\textfont2
        A\kern-.1667em\lower.5ex\hbox{M}\kern-.125emS}}
\def\AmSLaTeX{\AmS-\LaTeX}
\def\BibTeX{\rm B{\sc ib}\TeX}
\twocolumn[\hsize\textwidth\columnwidth\hsize\csname@twocolumnfalse%
\endcsname
\title{Long-range correlations and generic scale invariance in classical 
                                                                       fluids\\
             and disordered electron systems}
\author{T.R.Kirkpatrick}
\address{Institute for Physical Science and Technology, and Department of Physics\\
University of Maryland,\\ 
College Park, MD 20742}
\author{D.Belitz}
\address{Department of Physics and Materials Science Institute\\
University of Oregon,\\
Eugene, OR 97403}

\date{\today}
\maketitle

\begin{abstract}
Long-ranged, or power-law, behavior of correlation functions in both space 
and time is discussed for classical systems and for quantum systems at
finite temperature, and is compared with the corresponding behavior in
quantum systems at zero temperature. The origin of the long-ranged
correlations is explained in terms of soft modes.
In general, correlations at zero temperature are of longer range than their
finite temperature or classical counterparts.
This phenomenon is due to additional soft modes that exist
at zero temperature.
\end{abstract}
]
\section{Introduction}
\label{sec:I}

Homogeneous functions, or power laws, of space and time do not contain
any intrinsic length or time scales, in contrast to, e.g., exponentials.
This so-called scale invariance is well known to occur at critical points, 
where the critical modes become soft, which leads to power-law correlation
functions.\cite{CritPhen}
Critical points are exceptional points in the phase diagrams
of materials, and reaching them requires the fine tuning of parameter
values. What is less well appreciated is the fact that many systems
display what is now known as generic scale invariance (GSI), that is,
power-law correlation functions in large regions of parameter space,
with no fine tuning required at all to see them. GSI is due to soft
modes that are not related to critical phenomena, but rather are due to
conservation laws, or are otherwise inherent to the system. In recent years
there has been a lot of attention paid to GSI, and the phenomenon has been
discussed in a wide variety of systems, ranging from classical fluids and 
liquid crystals to disordered electrons and sandpiles. For recent reviews 
see Refs.\ \onlinecite{Reviews}.

In this paper we compare and contrast GSI and related phenomena in classical
fluids and Lorentz gases with what we will call quantum GSI (QGSI), a
similar but in general stronger effect that
occurs in quantum systems at zero temperature. We will discuss four
closely related topics: (1) The non-existence of virial expansions for
the transport coefficients, (2) long-time tails, i.e. power-law decays of
equilibrium time correlation functions that determine the transport
coefficients, (3) long-range, or power-law, spatial correlations in
non-equilibrium steady states, and (4) power-law spatial correlations
in quantum mechanical systems in and out of equilibrium.

In Sec.\ \ref{sec:II} of this paper we review the occurrence of the above 
phenomena in classical fluids and Lorentz gases. In Sec.\ \ref{sec:III}
we discuss the analogous effects in a quantum system, namely disordered 
electrons. In general, QGSI is characterized by correlations of longer 
range in space and/or time than those characteristic of the classical GSI.
Reasons for this are discussed. In Sec.\ \ref{sec:IV} we conclude with
a summary and a few remarks.

\section{Review of Classical Systems}
\label{sec:II}

\subsection{Density expansion of the transport coefficients}
\label{subsec:II.A}

The first indications for long-ranged dynamical correlations in classical
fluids appeared in the 1960's, when problems were encountered in attempts
to theoretically establish the density dependence 
of the transport coefficients of
moderately dense gases. Up to that time it had been assumed that, in
analogy with the virial expansion for the thermodynamic quantities\cite{Hill}
such as, e.g., the pressure, an analytic density expansion existed for an
arbitrary transport coefficient, $\eta$, of the form,
\begin{equation}
\eta/\eta_B = 1 + a_{1,\eta}\,n^* + a_{2,\eta}\,(n^*)^2 + O\left((n^*)^3\right)
                                                                      \quad.
\label{eq:2.1}
\end{equation}
Here $\eta_B$ is the Boltzmann value for $\eta$, which is exact in the limit
of vanishing fluid density, $n\rightarrow 0$. $n^* = n\sigma^d$ is the 
reduced density, with $\sigma$ a molecular length scale and $d$ the spatial
dimensionality of the system. The numbers $a_{l,\eta}$ are the coefficients
in the virial expansion of $\eta$.

A density expansion of the form of Eq.\ (\ref{eq:2.1}) was predicted by
Bogoliubov's theory,\cite{Bogoliubov}
an extension and generalization of Boltzmann's
kinetic theory to higher densities. This theory was generally accepted
at the time, and it gave the coefficients $a_{l,\eta}$ in
terms of formally exact integrals. It came as a big surprise when a 
detailed analysis of these integrals showed that
$a_{l,\eta}$ diverges for $l\geq 1$ in two-dimensional ($2$-$d$) systems, 
and for $l\geq 2$ in $3$-$d$ systems.\cite{DorfmanCohenWeinstock}
Shortly after this discovery, exact
calculations for a model fluid, the classical Lorentz gas, confirmed these
results.\cite{LorentzModel}
A Lorentz gas consists of noninteracting particles that move in
an array of randomly positioned, static hard disk ($d=2$) or hard sphere 
($d=3$) scatterers.\cite{Hauge}
The moving particles can be either classical or quantum
in nature. The only relevant transport coefficient is the diffusion
coefficient, or, equivalently, the particle mobility or conductivity. The
important point was that the exact explicit calculations for classical
Lorentz gases showed that the coefficients $a_{l,D}$ in the virial
expansion, Eq.\ (\ref{eq:2.1}), for the diffusion coefficient are indeed
divergent for $l\geq l^*$, and that the value of $l^*$ displays the 
dimensionality dependence suggested by the estimates for real gases. For
later reference, we mention that the quantum version of the Lorentz gas
is essentially the standard Edwards model for systems of disordered, 
noninteracting electrons.\cite{Edwards}

These divergencies in the virial coefficients are a clear indication of
the presence of long-ranged dynamical correlations in both Lorentz gases
and real gases. They have the following origin: In making a virial
expansion, one assumes that the $l$-th term in Eq.\ (\ref{eq:2.1}) is
determined by the dynamics of a cluster of $l+2$ particles moving in the
infinite system. Because the cluster is considered in isolation, the
$l+2$ particles can travel over arbitrarily large distances between 
collisions. However, this can occur neither in a real gas nor in a Lorentz 
gas, since the presence of the other particles that are not members of the
cluster under consideration means that a particle cannot travel farther than 
a distance on the
order of a mean-free path before it collides with another particle.
This is a collective many-particle effect that was missed in Bogoliubov's
cluster expansion. Its existence precludes a virial expansion for
transport coefficients. Mathematically, this physical effect leads to a
nonanalytic density dependence of the transport coefficients, i.e. the
virial coefficients are not simply numbers, but nonanalytic functions of
the density. The leading nonanalyticity turns out to be 
logarithmic,\cite{Sengers} in
agreement with the logarithmic divergence found in the early work.
Equation\ (\ref{eq:2.1}) then takes the form, in $d=3$,
\begin{eqnarray}
\eta/\eta_B = 1 + a_{1,\eta}\,n^* + {a'}_{2,\eta}\,(n^*)^2 \log n^* 
              + a_{2,\eta}\,(n^*)^2 
\nonumber\\
+\ o\left((n^*)^2\right)\quad.
\label{eq:2.2}
\end{eqnarray}
Here $o\left((n^*)^2\right)$ denotes a term that for $n^*\rightarrow 0$
vanishes faster than $(n^*)^2$.
For real gases, only estimates of the coefficient ${a'}_{2,\eta}$ are
known.\cite{LawSengers}
For the classical Lorentz gas, ${a'}_{2,D}$ is known exactly.\cite{LorentzModel}

\subsection{Long-time tails}
\label{subsec:II.B}

Up until the late 1960's it was also thought that the equilibrium time
correlation functions that determine the transport coefficients decay
exponentially with time for long times. This belief was based again on
Bogoliubov's kinetic theory, which predicted such an exponential decay.
Actually, the separation of time scales that is present only with
exponentially decaying time correlations was an important presumption
in that theory. This appeared to be very plausible since the Boltzmann 
equation, 
which becomes exact in the limit of dilute gases, also yields an exponential
decay of time correlations, which seemed to guarantee an exponential decay
at least for dilute gases. It was therefore a further completely unexpected
development when Alder and Wainwright,\cite{AlderWainwright}
in a computer study of self-diffusion 
in systems of hard disks and hard spheres, discovered that the equilibrium 
velocity autocorrelation function, 
$\langle{\bf v}(0)\cdot {\bf v}(t)\rangle_{\rm eq}$, 
whose time integral determines the diffusion coefficient, $D$, via
\begin{equation}
D = {1\over d} \int_{0}^{\infty} dt\ \langle{\bf v}(0)\cdot 
                                           {\bf v}(t)\rangle_{\rm eq}\quad,
\label{eq:2.3}
\end{equation}
decays only algebraically with time, namely,
\begin{equation}
\langle{\bf v}(0)\cdot {\bf v}(t)\rangle_{\rm eq} \approx c\,(t_0/t)^{d/2}\quad
                                 ({\rm for}\quad t>>t_0)\quad,
\label{eq:2.4}
\end{equation}
where $t_0$ is the mean-free time between collisions. Here $<\ldots >_{\rm eq}$
denotes an equilibrium thermal average.
This slow decay of correlation functions is known as a long-time tail (LTT).
The constant $c$ is positive, which implies that the LTT contribution to
the autocorrelation function increases the diffusion rate compared to
the Boltzmann result. Note that Eqs.\ (\ref{eq:2.3}) and (\ref{eq:2.4})
imply that the diffusion coefficient $D$ does not exist for $d\leq 2$,
and that for low frequencies, the frequency dependent diffusion 
coefficient, $D(\omega)$, behaves as,
\begin{mathletters}
\label{eqs:2.5}
\begin{equation}
D(\omega)/D_0 = 1 - c'\,(i\omega)^{(d-2)/2} + \ldots \quad (d>2)\quad,
\label{eq:2.5a}
\end{equation}
\begin{equation}
D(\omega)/D_0 = 1 - c''\,\log (i\omega) + \ldots \quad (d=2) \quad,
\label{eq:2.5b}
\end{equation}
\end{mathletters}%
with $D_0$ the static or zero-frequency diffusion coefficient, and
$c'$ and $c''$ positive constants.

Equation\ (\ref{eq:2.4}) was later derived theoretically by Ernst, Hauge,
and van Leeuwen,\cite{ErnstHaugevanLeeuwen} and by Dorfman and 
Cohen.\cite{DorfmanCohen} The basic physical idea behind the explanation
of the LTT phenomena is that it is the hydrodynamic processes 
that determine the
long-time behavior of all correlation functions. Of particular importance
are recollision processes, where after a collision, the two involved
particles diffuse away and then come back and recollide. We can see this in
Eq.\ (\ref{eq:2.4}), the right-hand side of which is proportional to the 
probability that a diffusing particle returns at time $t$ to the point
it started out from at $t=0$. This concept is a very general one, and it 
turns out that {\it all} of the transport coefficients
in a classical fluid have LTT like those shown in Eqs.\ (\ref{eq:2.4}) and
(\ref{eqs:2.5}). That is, if $\eta(\omega)$ is a general frequency dependent
transport coefficient, then
\begin{mathletters}
\label{eqs:2.6}
\begin{equation}
\eta (\omega)/{\eta}_0 = 1 - c_{\eta}'\,(i\omega)^{(d-2)/2} + \ldots \quad (d>2)\quad,
\label{eq:2.6a}
\end{equation}
\begin{equation}
\eta (\omega)/{\eta}_0=1 - c_{\eta}''\,\log (i\omega) + \ldots \quad (d=2) \quad.
\label{eq:2.6b}
\end{equation}
\end{mathletters}%

The existence of LTT even for arbitrarily dilute fluids is not in contradiction
with the Boltzmann equation becoming exact in the dilute limit. The point is
that the Boltzmann equation becomes exact for fixed time in the limit of
vanishing density, but {\it not} for fixed density, no matter how small, in the
limit of long times. The way the dilute limit is reached is that with
decreasing density, one has to go to longer and longer times in order to
see the LTT, and the preasymptotic decay is well described by the
Boltzmann equation.

It was pointed out above that the LTT in a fluid are related to the probability
of a diffusing particle to return at time $t$ to the point where it was at
time $t=0$. Although the recollision events that are responsible for this
return probability always occur, this does not necessarily imply 
that all correlation functions in a given system (other than a real
classical fluid) decay as
$t^{-d/2}$. Rather, it just suggests the possibility that they do, and
whether or not a particular correlation function actually does so, depends
on whether the corresponding observable couples sufficiently strongly to
the hydrodynamic processes in the system under consideration. 
It is plausible to assume that, for any given observable,
coupling to the diffusion process is more likely the more hydrodynamic or
soft modes there are in the system. For example, in a $d$-dimensional real
fluid there are $d+2$ soft modes due to the $d+2$ conservation laws for
particle number, energy, and momentum. In contrast, in a Lorentz gas 
only particle number (and, trivially, energy) is conserved. This smaller
number of hydrodynamic modes suggests, according to the above argument,
that the LTT might be weaker in a Lorentz gas than in a real fluid. Indeed,
the velocity autocorrelation function  in a Lorentz gas decays 
as,\cite{LorentzLTT}
\begin{equation}
\langle{\bf v}(0)\cdot {\bf v}(t)\rangle_{\rm eq}\approx 
       -c\,(t_0/t)^{(d+2)/2}\quad ({\rm for}\quad t>>t_0)\quad,
\label{eq:2.7}
\end{equation}
and the frequency dependent diffusion coefficient is given by
\begin{mathletters}
\label{eqs:2.8}
\begin{equation}
D(\omega)/D_0 = 1 + a\,i\omega + b\,(i\omega)^{d/2} + \ldots \quad (d>2)\quad,
\label{eq:2.8a}
\end{equation}
\begin{equation}
D(\omega)/D_0 = 1 + b'\,i\omega\,\log (i\omega) + \ldots \quad (d=2) \quad.
\label{eq:2.8b}
\end{equation}
\end{mathletters}%

The coefficients $c$, $b$, and $b'$ in Eqs.\ (\ref{eq:2.7}) and (\ref{eqs:2.8})
are positive. This sign occurs because in a Lorentz gas, in contrast to
a real fluid, any recollision process decreases the diffusion rate. This
is a consequence of the missing dynamics of the scatterers.

\subsection{Long-range correlations in nonequilibrium steady states}
\label{subsec:II.C}

In contrast to the slow time decay of equilibrium time correlation functions
for classical fluids discussed in the last subsection, the {\it spatial}
correlations decay exponentially in such systems, except at isolated
critical points. This asymmetry between space and time correlations
is related to the detailed balance relation that is valid in
equilibrium, and it is not generic. In more general states, e.g. in
nonequilibrium steady states, where detailed balance is absent
and where there is spatial anisotropy, long-range correlations occur in
both space and time.\cite{NESS}

From a general point of view the existence of long-ranged spatial correlations
is not surprising. Thermal fluctuations constantly appear in a fluid, and
then decay. Their behavior at long distances is determined by the soft
dynamical modes that are singular in the long-wavelength limit. For example,
in the static or zero-frequency limit, the diffusion equation becomes
Laplace's equation whose solution exhibits a power-law decay in space.

The best studied system, both theoretically and experimentally, is a
simple fluid subject to a stationary temperature 
gradient.\cite{NESS,GradTFluid} Using light
scattering, the Fourier transform of the density autocorrelation function,
\begin{equation}
S({\bf x},{\bf x}') = \langle\delta n({\bf x})\ \delta n({\bf x}')\rangle
                                            \quad,
\label{eq:2.9}
\end{equation}
can be measured. Here $\delta n({\bf x})$ is a density fluctuation at point
${\bf x}$, and the angular brackets denote a nonequilibrium ensemble
average. Theoretically, both microscopic many-body techniques and more
phenomenological approaches yield,\cite{KCD}
\begin{equation}
S({\bf k}) = S_0\,\left[1 + \left({{\hat {\bf k}}_{\perp}\cdot\nabla T
          \over {\bf k}^2}\right)^2\ {c_p\over T\,D_T(\nu + D_T)}\right]\quad.
\label{eq:2.10}
\end{equation}
Here ${\bf k}$ is the wavevector, $c_p$ is the specific heat at constant
pressure, $D_T$ is the thermal diffusion coefficient, $\nu$ is the
kinematic viscosity, ${\hat{\bf k}}_{\perp}$ is a unit vector
perpendicular to ${\bf k}$, and $S_0$ is the equilibrium static structure
factor that is $k$-independent in the long wavelength limit. Equation\ 
(\ref{eq:2.10}) for $S({\bf k})$ is well confirmed experimentally.
For the implications of Eq.\ (\ref{eq:2.10}) in real
space, see Ref.\ \onlinecite{Schmitz}.

Theoretically, long-range correlations in a nonequilibrium Lorentz gas 
have also been studied. The system considered is a Lorentz gas in the
presence of a chemical potential gradient, $\nabla\mu$. As mentioned
earlier, the only soft mode in this model is the diffusive number density.
Because the diffusing particles in a Lorentz model
do not interact, correlations can only occur between the moving particles
and the scatterers. Denoting the density fluctuations of the former by
$\delta n({\bf x})$, and the scatterer density for a given configuration
by $N({\bf x})$, a measure of this correlation is the particle-scatterer
density correlation,
\begin{equation}
S({\bf x},{\bf x}') = \langle \delta n({\bf x})\,N({\bf x}')\rangle\quad.
\label{eq:2.11}
\end{equation}
In Fourier space, a kinetic theory calculation yields,\cite{NonEqLorentz}
\begin{equation}
S({\bf k}) = {i{\bf k}\cdot\nabla\mu\over 2\pi D {\bf k}^2}\quad,
\label{eq:2.12}
\end{equation}
with $D$ the diffusion coefficient. In real space, Eq.\ (\ref{eq:2.12})
implies that $S(r)$ decays like $r^{1-d}$. Again, the correlations are
stronger in the real fluid than in the Lorentz gas.

\section{Disordered Electrons at Zero Temperature}
\label{sec:III}

\subsection{Density expansion of the electron mobility}
\label{subsec:III.A}

In disordered electronic systems, there are two length scales that
can be used to construct a dimensionless small parameter from the
scatterer density $n$, namely the scattering length $\sigma$ and
the Fermi wavelength $\lambda_F = 2\pi/k_F$. We can thus form the
dimensionless densities (for $d=3$) $n\lambda_F^3$, $n\lambda_F^2\sigma$,
$n\lambda_F\sigma^2$, and $n\sigma^3$. The first two do not appear in the
standard perturbation theory that expands in powers of $n$.\cite{Edwards}
The third one is usually written as $1/k_Fl\equiv\epsilon$, with 
$l\sim 1/n\sigma^2$ the mean-free
path. The fourth one is the quantity $n^* = n\sigma^3$ that also has a
classical meaning. Now consider a dilute electron system in the sense
that $\lambda_F >> \sigma$. In that case we have 
$n^* \sim \sigma/l << \epsilon$, and hence $n^*$ can be neglected
compared to $\epsilon$. Notice that in the quantum Lorentz gas, the
dilute limit needs to be considered even for noninteracting quantum
particles, since the Pauli principle establishes correlations,
and hence an effective interaction, between the particles.

For dimensions $d>2$, the system is diffusive as long as 
$\epsilon<\epsilon^*$, with $\epsilon^* = O(1)$ for $d=3$. At
$\epsilon = \epsilon^*$ a metal-insulator transition called the Anderson
transition takes place.\cite{LeeRama}
For dilute systems in the above sense, $n^*<<1$
even at the Anderson transition, and the latter is well separated from the
percolation transition that would occur at $n^*\approx 1$.

Considering the generalized virial expansion given by Eq.\ (\ref{eq:2.2})
it is natural to ask whether or not a similar expression holds for a
quantum Lorentz gas with the parameter $\epsilon$ defined above replacing
the classical reduced density $n^*$. This question was first asked, and
answered affirmatively, as early as 1966.\cite{LangerNeal} 
More recently, all of the
terms in the expansion up to and including those of $O(\epsilon^2)$ have
been computed exactly.\cite{Wyso}
In order to compare the resulting expression with
real experimental data, the electron mobility, $\mu$, of electrons injected
into Helium gas at low temperatures has been considered. Because the
Helium atoms are very massive compared to the electrons, the quantum Lorentz
gas constitutes an excellent model for this system. Furthermore, there is
experimental control over the density of the injected electrons, which means
that the abovementioned diluteness condition can be fulfilled to an extremely
high degree. This also means that Coulomb interaction effects can be 
made negligible. In order to model the
actual experimental situation, one needs to consider a nondegenerate gas
of electrons at temperature $T$ with the thermal de Broglie wavelength, 
$\lambda_T = 2\pi/k_T = (2\pi/m k_B T)^{1/2}$, 
replacing $\lambda_F$ in $\epsilon$.
Defining $\chi = 2/k_T l$, an exact calculation 
gives,\cite{Wyso}
\begin{mathletters}
\label{eqs:3.1}
\begin{equation}
\mu/\mu_{\rm cl} = 1 + \mu_1\,\chi
              + {\mu}'_{2}\,\chi^2 \log\chi
              + \mu_2\,\chi^2
              + o\left(\chi^2\right) \quad,
\label{eq:3.1a}
\end{equation}
with
\begin{eqnarray}
\mu_1 = -\pi^{3/2}/6\quad\quad,
\nonumber\\
{\mu}'_{2} = \left(\pi^2 - 4\right)/32\quad,
\label{eq:3.1b}\\
\mu_2 = 0.236\ldots\quad\quad,
\nonumber
\end{eqnarray}
\end{mathletters}%
and $\mu_{\rm cl}$ the classical or Boltzmann value for the mobility.

\begin{figure}
\epsfxsize=8.25cm
\epsfysize=5.8cm
\epsffile{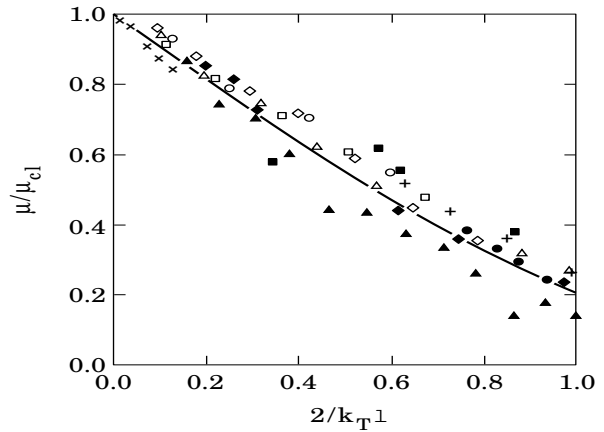}
\vskip 0.5cm
\caption{Mobility $\mu$ of electrons in dense gases, normalized to the
 Boltzmann value $\mu_{\rm cl}$, as a function of $\chi$. The symbols
 represent experimental data, the solid line is the theoretical result,
 Eq.\ (\protect\ref{eq:3.1a}). After Ref.\ \protect\onlinecite{Adams}.}
\label{fig:1a}
\end{figure}

\begin{figure}
\epsfxsize=8.25cm
\epsfysize=5.5cm
\epsffile{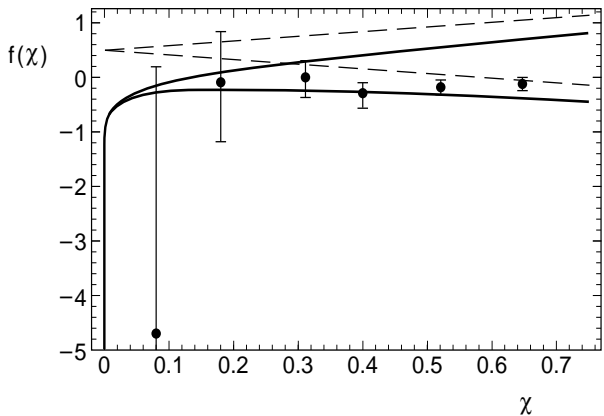}
\vskip 0.5cm
\caption{The reduced mobility $f$, as defined in
 Eq.\ (\protect\ref{eq:3.2a}),
 vs. the density parameter $\chi=2/k_T l$. The theoretical
 prediction is for $f$ to lie between the two solid lines. The
 experimental data are from Fig. 9 of Ref.
 \protect\onlinecite{Schwarz} with
 error bars estimated as described in Ref.\ \protect\onlinecite{Wyso}.
 The broken lines
 show what the theoretical prediction would be in the absence of
 the logarithmic term in the density expansion. 
 From Ref.\ \protect\onlinecite{Wyso}.}
\label{fig:1b}
\end{figure}

In Figs.\ \ref{fig:1a} and \ref{fig:1b} this theoretical result is
compared with experimental data. Of particular
interest is the question whether this kind of analysis can be used
to experimentally confirm the existence of the logarithmic term in
Eq.\ (\ref{eq:3.1a}). In classical systems the corresponding
logarithmic term in Eq.\ (\ref{eq:2.2}) has never been convincingly
observed,\cite{LawSengers}
mostly due to the fact that the coefficients in the density
expansion are not known exactly for any realistic classical system. 
The fact that the quantum Lorentz gas is such a good model for electrons
in Helium gas makes this system a very promising one for attempts to
finally observe the logarithm. As a measure of the logarithmic term,
one defines,\cite{Wyso}
\begin{mathletters}
\label{eqs:3.2}
\begin{equation}
f(\chi) \equiv \left[\ \mu/\mu_{\rm cl} - 1 - \mu_1\,\chi\ \right]
                                 /\chi^2 \quad.
\label{eq:3.2a}
\end{equation}
The theoretical prediction for this quantity is,
\begin{equation}
f(\chi) = {\mu}'_{2}\,\log\chi + \mu_2 \pm\mu_2\,2\sqrt{\pi}
                                                    \chi\quad.
\label{eq:3.2b}
\end{equation}
\end{mathletters}%

\noindent
The last term in Eq.\ (\ref{eq:3.2b}) is
an estimate of the effect of all higher order terms.
At $T=4.2{\rm K}$, a Helium gas density of $n=10^{21}{\rm cm}^{-3}$
corresponds to $\chi = 1$, and data were obtained for $\chi$
as small as $0.08$. Fig.\ \ref{fig:1b} shows the theoretical prediction,
Eq.\ (\ref{eq:3.2b}), for $0 < \chi < 0.7$ together with data by
Schwarz. We conclude that the existing experimental data are consistent with
the theoretical result. However, for a convincing demonstration
of the existence of the logarithmic term an improvement in the 
experimental accuracy by about a
factor of ten over Schwarz's experiment would be necessary.

\subsection{Long-time tails, a.k.a. weak localization effects}
\label{subsec:III.B}

The results discussed in the previous subsection seems to suggest that there is
no conceptual difference between transport in classical and dilute quantum
Lorentz gases: The forms of the density or disorder expansions,
Eqs.\ (\ref{eq:2.2}) and (\ref{eqs:3.1}), are identical, even though the
dimensionless expansion parameters are different. This conclusion is
fallacious, however, as can be seen by considering the LTT in the time
correlation functions, or in the low frequency expansions of the
transport coefficients for the two models. Here we first quote the
quantum result, and then we discuss the reason for it being qualitatively
different from its classical counterpart.

\begin{figure}
\epsfxsize=8.25cm
\epsfysize=5.8cm
\epsffile{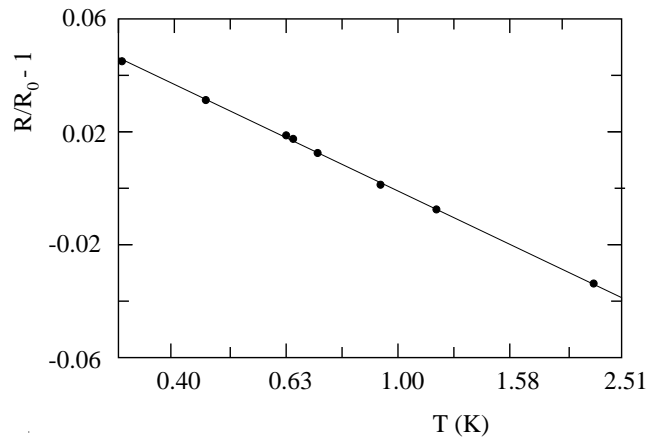}
\vskip 0.5cm
\caption{Resistance, $R$, normalized to $R_0 = R(T=1K)$,
 of a thin PdAu film plotted versus $\log T$. After
 Ref.\ \protect\onlinecite{DolanOsheroff}.}
\label{fig:2}
\end{figure}

Any of a variety of theoretical methods leads to 
a frequency dependent electrical
conductivity for a quantum Lorentz gas whose real part is of the 
form,\cite{GLK,KD}
\begin{mathletters}
\label{eqs:3.3}
\begin{equation}
\sigma(\omega)/\sigma_0 = 1 + c\,\omega^{(d-2)/2} + 
                                     \ldots \quad ({\rm for}\quad d>2) \quad,
\label{eq:3.3a}
\end{equation}
\begin{equation}
\sigma(\omega)/\sigma_0 = 1 + c'\,\log\omega + \ldots \quad ({\rm for}\quad d=2)
                                                                   \quad.
\label{eq:3.3b}
\end{equation}
Equations\ (\ref{eq:3.3a}), (\ref{eq:3.3b}) imply that the current-current
correlation function that is defined as the Fourier transform of the real
part of the conductivity has a LTT,
\begin{equation}
\sigma(t) \approx -c\,(t_0/t)^{d/2}\quad ({\rm for}\quad t>>t_0)\quad.
\label{eq:3.3c}
\end{equation}
\end{mathletters}%
The coefficient $c$ in Eq.\ (\ref{eq:3.3c}) is positive. For
$d=2$, and more generally for $d\leq 2$, the low frequency expansion of
$\sigma(\omega)$ breaks down, and the static conductivity or the static 
diffusion coefficient is actually zero. That is, for $d\leq 2$ there is no
metallic phase at zero temperature.\cite{g4,LeeRama}
At finite temperature and zero
frequency, the temperature dependence of $\sigma$ is obtained by replacing
the frequency in Eqs.\ (\ref{eq:3.3a}) and (\ref{eq:3.3b}) by the
temperature $T$. For $T>0$ and $\omega\rightarrow 0$ the leading frequency
dependences are given by Eqs.\ (\ref{eqs:2.8}), i.e., the classical
result is recovered. Figure\ \ref{fig:2} shows an example of the temperature
dependent resistivity of a thin metallic film, which is logarithmic for
low temperatures in agreement with the above remarks.

Let us now discuss the interesting difference between the frequency
dependencies in Eqs.\ (\ref{eqs:2.8}) and (\ref{eqs:3.3}). Equations\ 
(\ref{eqs:2.8}) can be derived not only from a classical microscopic
many-body approach, but also from a more general phenomenological approach
that appears to be independent of whether or not the underlying
description is classical or quantum mechanical.\cite{ErnstMachta}
The crucial assumption is
that the only slow mode in the problem is due to particle number
conservation. Remarkably, it is this assumption that breaks down in the
quantum case, and this
is what leads to the differences between Eqs.\ (\ref{eqs:2.8}) and
(\ref{eqs:3.3}). To understand this important point, let us consider
a field theoretic description of a disordered fermion system,\cite{NO} that we
assume to be noninteracting for simplicity. The partition function is,
\begin{mathletters}
\label{eqs:3.4}
\begin{equation}
Z = \int_{\psi(0)=-\psi(1/k_B T)} D[\bar\psi,\psi]\ 
                                  \exp\left(S[\bar\psi,\psi]\right)\quad,
\label{eq:3.4a}
\end{equation}
with the action $S$ given by,
\begin{eqnarray}
S[\bar\psi,\psi] = -\int dx\ \bar\psi(x)\,{\partial\over\partial\tau}\,
                                                  \psi(x) + 
\nonumber\\
                   + \int dx\ \bar\psi(x)\,\left[{\Delta\over 2m} + \mu
                          - u({\bf x})\right]\,\psi(x)\quad.
\label{eq:3.4b}
\end{eqnarray}
\end{mathletters}%
Here we have used a four-vector notation, $x\equiv ({\bf x},\tau)$,
$\int dx \equiv \int d{\bf x}\,\int_0^{1/T} d\tau$, with $\tau$ denoting
imaginary time. $m$ is the fermion mass, $\mu$ is the chemical potential,
$u({\bf x})$ is a random
potential, and for notational simplicity we have suppressed the spin
labels. Since we are considering fermions, the fields $\bar\psi$ and
$\psi$ are Grassmann valued and $D[\bar\psi,\psi]$ is a Grassmannian
functional integration measure, but for the following arguments this will
not be crucial. By changing from imaginary time representation to a
frequency representation,
\begin{mathletters}
\label{eqs:3.5}
\begin{equation}
\psi(x) = T^{1/2} \sum_n e^{-i\omega_n\tau}\,\psi_n({\bf x})\quad,
\label{eq:3.5a}
\end{equation}
with Matsubara frequencies
\begin{equation}
\omega_n = 2\pi T (n+1/2)\quad,\quad (n=0,\pm 1,\ldots)\quad,
\label{eq:3.5b}
\end{equation}
\end{mathletters}%
the action can be written,
\begin{equation}
S = \sum_n \int d{\bf x}\ \bar\psi_n({\bf x})\,\left[{1\over 2m}\Delta + \mu
                 - u({\bf x}) - i\omega_n\right]\,\psi_n({\bf x})\quad.
\label{eq:3.6}
\end{equation}
The crucial point is that for $\omega_n = 0$, or $T=0$, the action given
by Eq.\ (\ref{eq:3.6}) is invariant under a unitary transformation of the
fields in frequency space, $\psi_n \rightarrow \sum_m U_{nm}\psi_m$. 
In fact, $S$ is invariant under a larger symplectic group that
also includes a time reversal symmetry, but we ignore this technical point
here. We further note that the `order parameter',
\begin{equation}
Q = \lim_{\omega_n\rightarrow 0+} \langle\bar\psi_n({\bf x})\,\psi_n({\bf x})
                                                                      \rangle
    - \lim_{\omega_n\rightarrow 0-} \langle\bar\psi_n({\bf x})\,\psi_n({\bf x})
                                                                 \rangle\quad,
\label{eq:3.7}
\end{equation}
is the single-particle spectral function, or the difference between the 
retarded and advanced Green functions. Because these functions have poles on
opposite sides of the real axis, $Q$ is nonzero as long as the density of
states at the Fermi surface is nonzero. 

To use a magnetic analogy, having a nonzero $Q$ in Eq.\ (\ref{eq:3.7}) is
similar to having a nonvanishing magnetization in the limit of a zero
external field, and the abovementioned unitary
symmetry is analogous to the rotational
symmetry in spin space. This analogy implies that in the zero temperature
fermion system there is a spontaneously broken continuous symmetry. This
was first noticed in the context of the Anderson transition mentioned in
Sec.\ \ref{subsec:III.A}.\cite{Wegner,McKaneStone} Goldstone's
theorem then implies that there are soft modes, namely particle-hole
excitations, in addition to the ones
implied by the conservation laws. Detailed calculations confirm that it is
these additional soft modes that lead to the stronger LTT effects in
Eqs.\ (\ref{eqs:3.3}) as compared to the classical LTT in 
Eqs.\ (\ref{eqs:2.8}).

\subsection{Long-ranged spatial correlations in equilibrium}
\label{subsec:III.C}

A characteristic feature of quantum statistical mechanics, as opposed to
the classical theory, is the coupling of statics and dynamics. This can
be seen in Eqs.\ ({\ref{eqs:3.4}), where the basic statistical field
$\psi$ is a function of both space and imaginary time. As a result, one
does not expect any qualitative differences between static and dynamic
correlations even in equilibrium, in contrast to the asymmetry between
these two types of correlations that is observed in classical systems and
was discussed in Sec.\ \ref{subsec:II.C} above.

Indeed, a calculation of the wavenumber dependent static spin susceptibility,
$\chi_s({\bf k})$, in a disordered system of interacting electrons yields 
the following behavior for small wavenumbers,\cite{fatpaper,fmlong}
\begin{equation}
\chi_s({\bf k}) = c_0 - c_{d-2}\,\vert {\bf k}\vert^{d-2} - c_2\,{\bf k}^2
                  + \ldots\quad,
\label{eq:3.8}
\end{equation}
where the $c_i$ are positive constants. In real space, the nonanalytic term
proportional to $\vert {\bf k}\vert^{d-2}$, which for $d<4$ is the leading
${\bf k}$-dependence of $\chi_s$, corresponds to a long-range interaction
between the electronic spin density fluctuations that falls off like
$r^{2-2d}$. This has recently been shown to have interesting consequences
for the ferromagnetic quantum phase transition that occurs in an
itinerant electron system at zero temperature as a function of the
exchange interaction.\cite{fmlong}

The origin of this long-range correlation can be traced back to the same
Goldstone modes that were discussed in the last subsection, and that
also lead to the LTT. In a disordered system, the Goldstone modes are
diffusive, and give a contribution to the spin susceptibility that can
be schematically represented by
\begin{equation}
\chi_s({\bf k}) \sim \int_k^{\Lambda} dp\ p^{d-1} \int_0^{\infty} d\omega
                  \ {\omega\over (D\,p^2 + \omega)^3}\quad,
\label{eq:3.9}
\end{equation}
with $\Lambda$ an ultraviolet cutoff, and $D$ the spin diffusion coefficient.
Equation\ ({\ref{eq:3.9}) demonstrates the coupling of statics and dynamics
that was mentioned above, and doing the integrals yields Eq.\ (\ref{eq:3.8}).
The fact that this coupling is really a quantum effect can be seen by
considering the corresponding expression at finite temperature. In this
case one has to perform a frequency sum rather than an integral, and the
net effect is that the term $\vert {\bf k}\vert^{d-2}$ in Eq.\ (\ref{eq:3.8})
is replaced by $({\bf k}^2 + T)^{(d-2)/2}$. Hence, for $T>0$ an analytic
expansion about ${\bf k}=0$ exists, and there are no long-ranged correlations.

\subsection{Long-ranged spatial correlations in nonequilibrium steady states}
\label{subsec:III.D}

Very recently, spatial correlations of density fluctuations have been 
studied in noninteracting disordered electronic system that are not in 
equilibrium.\cite{Yoshi2}
For the model defined by Eqs.\ (\ref{eqs:3.4}) or (\ref{eq:3.6}) the
correlation function analogous to Eq.\ (\ref{eq:2.11}) for the classical
Lorentz gas is,
\begin{equation}
S_1({\bf x},{\bf x}') = \left\{\langle\delta n({\bf x})\rangle\,u({\bf x}')
                                      \right\}_{\rm dis}\quad,
\label{eq:3.10}
\end{equation}
where $\{\ldots\}_{\rm dis}$ denotes the disorder average, and $<\ldots >$
denotes a nonequilibrium thermal average as in Sec.\ \ref{subsec:II.C}.
A direct many-body calculation shows that the Fourier transform of
Eq.\ (\ref{eq:3.10}), $S_1({\bf k})$, behaves just like its classical
counterpart, Eq.\ (\ref{eq:2.12}).

Because the moving particles are fermions, they effectively interact due
to statistical correlations. As a measure of these correlations we consider
the structure factor,
\begin{equation}
S_2({\bf x},{\bf x}') = \left\{\langle\delta n({\bf x})\,\delta n({\bf x}')
                                            \rangle\right\}_{\rm dis}\quad.
\label{eq:3.11}
\end{equation}
Let us consider the nonequilibrium part of $S_2$.
For a classical, interacting, Lorentz gas one finds,
\begin{equation}
S_2({\bf k}) \sim (\nabla\mu)^2/{\bf k}^2\quad.
\label{eq:3.12}
\end{equation}
Naively, one might anticipate a similar result for the disordered electron
system at $T=0$. However, because of the additional soft modes that were
discussed in the preceding section, the correlations here are much stronger,
and the decay is much slower in space. A direct many-body calculation
yields in the limit of small wavenumbers,
\begin{equation}
S_2({\bf k}) = {N_F\mu\tau\over 6d\pi (D{\bf k}^2)^2}\,\left[
                  25(\nabla\mu)^2 - 12({\hat{\bf k}}\cdot\nabla\mu)^2\right]
                                                                    \quad,
\label{eq:3.13}
\end{equation}
with $N_f$ the electronic density of states at the Fermi level, and $\tau$
the electronic mean-free time between collisions. In real space, 
Eq.\ (\ref{eq:3.13}) corresponds to a linear decay of 
$S_2(r)$ with distance.

\section{Conclusion}
\label{sec:IV}

In this paper we have reviewed classical and quantum versions of what one
might call generalized long-time tail effects, that is long-range correlations
in both space and time. We have seen that these effects are due to the
hydrodynamic or soft modes in the system, which couple via mode-mode
coupling effects to all other modes unless some symmetry prohibits such a
coupling. Generally, the quantum or zero temperature versions
of these effects are stronger, that is of longer range, than their classical
counterparts, because of additional soft modes that exist at zero
temperature. These additional soft modes are Goldstone modes that result
from a broken symmetry in Matsubara frequency space, and are not related
to conservation laws. Another additional effect in quantum systems is the
coupling of statics and dynamics, which leads to both static and dynamic
equilibrium correlations in general to be of long range, whereas in
classical systems one has to consider nonequilibrium states in order to
get long-ranged static correlations.

A phenomenon similar to the enhanced long-time tail effects in the quantum
case is also known in certain classical systems with soft modes that are
unrelated to conservation laws. For example, in the smectic-A phase
of liquid crystals there are soft modes due to the conservation
laws, and additional soft modes due to spontaneously broken symmetries
and Goldstone's theorem. The combination of these soft modes produce
stronger long-time tail effects than are present in a simple classical
fluid with no Goldstone modes.\cite{MRT}

Finally, we mention that the effects in electronic systems discussed in
Sec.\ \ref{sec:III} do not qualitatively hinge on the system being
disordered. As can be seen from Eq.\ (\ref{eq:3.6}) and the related
discussion, the broken symmetry argument still holds for clean electron
fluids, with the only difference being that the Goldstone modes have a 
ballistic dispersion in
that case, rather than a diffusive one. Consequently, the effects
discussed in this paper qualitatively survive, only the various exponents
change compared to the disordered system. These LTT effects in clean
fermion systems can be related to known features of Fermi-liquid theory.
This demonstrates the generality and unifying properties of the general
physical approach taken in this paper.

\acknowledgments

It is our pleasure to dedicate this paper to Matthieu H. Ernst on the
occasion of his sixtieth birthday. Matthieu has been one of the pioneers
in discovering and understanding the physical phenomena discussed above.
His early work on classical systems laid much of the basis for later
developments, and he has remained at the forefront of research in this
field.

This work was supported by the National Science Foundation under Grant Nos.
DMR-92-17496 and DMR-95-10185.

\end{document}